\documentclass[12pt]{article}
\usepackage[symbol]{footmisc}
\def\correspondingauthor{\footnote{Corresponding author.  }}
\usepackage[left=2.5cm,top=2.5cm,right=2cm,bottom=2.5cm]{geometry}
\usepackage{amsmath, amssymb}

\usepackage{graphicx}
\usepackage{graphics}
\usepackage{epstopdf}
\usepackage{subfigure}
\usepackage{amsfonts}
\usepackage{sectsty}
\usepackage{sectsty}
\usepackage{hyperref,lineno}
\usepackage{cite}

\begin{document}
	
	\begin{center}
	\large{\bf{
	Cosmological dynamics in $R^2$ gravity with logarithmic trace term}
	} 	\\

	\vspace{4mm}
	\normalsize{Emilio Elizalde$^{1}$, Nisha Godani$^2$ and Gauranga C. Samanta$^{3} {}$\correspondingauthor{} }\\
	\normalsize{$^1$Institute of Space Sciences (IEEC-CSIC), Campus UAB,\\
Carrer de Can Magrans, s/n, 08193 Cer,danyola del Valles, Barcelona, Spain\\
$^2$Department of Mathematics, Institute of Applied Sciences and Humanities\\ GLA University, Mathura, Uttar Pradesh, India\\
$^{3}$Department of Mathematics, BITS Pilani K K Birla Goa Campus, Goa, India}\\
	\normalsize {elizalde@ieec.uab.es; nishagodani.dei@gmail.com; gauranga81@gmail.com}
\end{center}
\begin{abstract}
A novel function for modified gravity is proposed, $f(R, T)=R+\lambda R^2+2\beta\ln(T)$, with constants $\lambda$ and $\beta$, scalar curvature $R$, and the trace of stress energy tensor $T$, satisfying $T=\rho-3p>0$. Subsequently, two equations of state (EoS) parameters, namely $\omega$ and a parametric form of the Hubble parameter $H$, are employed in order to study the accelerated expansion and initial cosmological bounce of the corresponding universe. Hubble telescope experimental data for redshift $z$ within the range $0.07\leq z \leq 2.34$ are used to compare the theoretical and observational values of the Hubble parameter. Moreover, it is observed that all the energy conditions are fulfilled within a neighborhood of the bouncing point $t=0$, what shows that the necessary condition for violation of the null energy condition, within a neighborhood of the bouncing point in general relativity, could be avoided by modifying the theory in a reasonable way. Furthermore, a large amount of negative pressure is found, which helps to understand the late time accelerated expansion phase of the universe.
\end{abstract}

\section{Introduction}
At the end of the past century, the accelerated expansion of our universe was discovered by  different cosmological surveys \cite{Riess, Perlmutter, Perlmutter1}. This finding was considered as one of the most important discoveries of the 20th century. Strenuous research has been going on since then, on this subject; however, the cause of this acceleration still remains unclear. New theories and modifications of the general theory of relativity have been suggested to explain the cosmic acceleration. In particular, two alternative possibilities have been intensively studied: either the universe contains an enormous amount of dark energy or the theory of general relativity breaks down at cosmological scales\cite{Frieman}.

The general theory of relativity can be modified in many different ways \cite{Tsujikawa, Capozziello,  Nojiri, Berti, Nojiri:2017ncd}. A comprehensive review on several modifications of general relativity and their cosmological consequences over the past few decades has been presented in \cite{Clifton}.
The scalar–tensor theories of gravity are  well studied modified theories of gravity, in the literature, which arise in the form of dimensionally reduced
effective theories of higher dimensional theories \cite{Bergmann, Nordtvedt}.
Fourth-order theories of gravity generalize the action with the replacement of the scalar curvature, $R$, in the Einstein gravitational
action by an arbitrary function, $f(R)$, with $R$ remaining the leading order contribution to $f(R)$ and are referred to as   $f(R)$ theories of gravity \cite{Schmidt,Sotiriou,Felice}.
 These  modified theories of gravity have gained  ample consideration for its capability to elucidate the accelerated expansion of the universe \cite{Buchdahl}. In the early 1980s,  Starobinsky \cite{Star} discussed a most simple $f(R)$ model, by taking $f(R)=R+\alpha R^2$, with $\alpha>0$, which is considered today as representing the first ever found inflationary scenario for the universe (and one of the most natural and successful).
The scenario to unify inflation with dark energy in a consistent way was proposed in \cite{Nojiri:2003ft}.
A simplification of $f(R)$ gravity suggested in \cite{Bertolami} integrates an unambiguous coupling between the matter Lagrangian and an arbitrary function of the scalar curvature, which leads to an extra force in the geodesic equation of a perfect fluid.
The description of cosmic acceleration by the modification of gravity at small curvature described by higher order differential equations may suffer violent instabilities \cite{Dolgov}.
 Subsequently, it is shown that this extra force may provide a justification for the accelerated expansion of the universe \cite{Nojiris, Bertolami1, Bamba}.
The dynamical behavior of the matter and dark energy effects have been obtained, within extended theories of gravity, in \cite{Nojiri001, Shirasaki, Capozziello001,  Rodrigues01}. Apart from that, recently  many authors have studied the dynamics of cosmological models in $f(R)$ gravity from various directions\cite{Capozziello91, Godani01, Bombacigno21, Sbis, Chen, Elizalde26, Elizalde25, Astashenok1, Miranda, Nascimento, Odintsov20, Elizalde:2004mq, Cognola:2006eg, Capozziello:2005pa, Elizalde:2008pv,  Odintsov1, Nojiri56, Parth, Samanta19, Godani19}.  Unsurprisingly, however, a viable $f(R)$ theory must fulfill very strict constraints. For example, solar system tests have ruled out a good deal of the $f(R)$ models suggested so far \cite{Chiba}. Nevertheless, a number of realistic consistent $f(R)$ gravity models which pass the
solar tests were proposed in \cite{Cognola:2007zu, Nojiri:2007cq, Nojiri21}.  \\

Further, modified Gauss-Bonnet (GB) gravity is an
another theory of gravitation which  is also called as $f(G)$ theory of gravity, where
$f(G)$ is a general function of GB invariant $G$ \cite{Nojiri2005}. A modification of GB theory is $f(G, T)$ gravity, which includes the matter contribution through the trace of the stress-energy tensor $T$ \cite{Sharif1}. This theory is used to find the exact solutions and compute the physical quantities in the context of an anisotropic model \cite{Shamir}. This theory has been investigated in several other aspects \cite{Sharif2, Shamir1, Sharif3}.
Apart from these, the class of $f(R, T)$ theories  have gained much attention to explain the accelerated expansion of the universe \cite{harko}.
In these theories the matter term is included in the gravitational action, in which the gravitational Lagrangian
density is an arbitrary function of both $R$ and $T$.
The random requirement on $T$ embodies the conceivable contributions from both non-minimal coupling and unambiguous $T$ terms.\\

Many functional forms of $f(R, T)$ models have been studied for the derivation of cosmological dynamics in different contexts.
The split-up case, $f(R, T)=f_1(R)+f_2(T)$, has received a lot of attention because it is simple and one can explore the contribution from
$R$ without specifying $f_2(T)$ and, similarly,  the contribution from $T$ without specifying $f_1(R)$. Reconstruction of $f(R, T)$ gravity in such separable theories is studied in \cite{Houndjo1}.
A non-equilibrium picture of thermodynamics at the apparent horizon of the (FLRW) universe was discussed in \cite{Sharif01}. Subsequently, many authors have been studying in depth this particular form of $f(R, T)=f_1(R)+f_2(T)$, to match different aspects of the cosmological dynamics with observations \cite{Jamil1, Alvarenga01, Santos1, Samanta, Shabani, Naidu, Samanta3, Chandel, Samanta1, Shabani001, Samanta2, Moraes01, Noureen, Farasat, Mirza, Correa1, Ramesh, Moraes0001, Moraes9, Zaregonbadi, Rahaman1, Mishra, Yousaf, Samanta5, Godani1, Elizalde01, Aditya, Elizalde, Ordines}. Bouncing cosmology in the framework of $f(R,T)$ gravity have been studied in \cite{Singh} by defining a particular form of the Hubble parameter in a flat, homogeneous and isotropic model.\\

The motivation of this paper is to study a  simple and very interesting form of  $f(R, T)$ gravity model, indeed a new functional form, defined as $f(R, T)=R+\lambda R^2+2\beta\ln(T)$, where $\lambda$, $\beta$ are constant and $T=\rho-3p>0$. From the above choice of $f(R, T)$ model,  the term $\rho-3p$ must be positive,  for the function $f(R, T)$ to be well defined. Therefore, the constraint $\rho-3p>0$ is mandatory. Subsequently, we will be able to affirm that  $\rho+3p>0$, provided $p>0$; now, if we are able to show in what follows that $\rho >0$, then all energy conditions, i.e., the Null Energy Condition (NEC), the Weak Energy Condition (WEC), and the Strong Energy Condition (SEC),  will be satisfied; and consequently, we will be able to assure that our model does not contain any exotic type of matter, for this particular choice of the $f(R, T)$ function.

In this paper, we use the natural system of units $G=c=1$.

\section{The background of $f(R, T)$ gravity}
The $f(R, T)$ theory was first introduced, in 2011, by Harko et al. \cite{harko}. These authors extended standard general theory of relativity by modifying the gravitational Lagrangian. The gravitational action in this theory is given by
\begin{equation}\label{action2}
S=S_G + S_m=\dfrac{1}{16\pi}\int f(R,T)\sqrt{-g}d^4x +\int \sqrt{-g}\mathcal{L} d^4x,
\end{equation}
where $f(R, T)$ is taken to be an arbitrary function of $R$ and $T$. Here, $R$ is the Ricci scalar and $T$ is the trace of the energy momentum tensor $T_{\mu\nu}$. The matter Lagrangian density is denoted by $\mathcal{L}$ and the energy momentum tensor is defined in terms of the matter action as follows \cite{Landau}:
\begin{equation}\label{}
  T_{\mu\nu}=-\frac{2\delta (\sqrt{-g}\mathcal{L})}{\sqrt{-g}\delta g^{\mu\nu}},
\end{equation}
which yields
\begin{equation}\label{}
  T_{\mu\nu}=g_{\mu\nu}\mathcal{L}-2\frac{\partial \mathcal{L}}{\partial g^{\mu\nu}}.
\end{equation}
The trace $T$ is defined as $T=g^{\mu\nu}T_{\mu\nu}$. Let us define the variation of $T$ with respect to the metric
tensor as
\begin{equation}\label{}
  \frac{\delta (g^{\alpha\beta}T_{\alpha\beta})}{\delta g^{\mu\nu}}=T_{\mu\nu}+\Theta_{\mu\nu},
\end{equation}
where $\Theta_{\mu\nu}=g^{\alpha\beta}\frac{\delta T_{\alpha\beta}}{\delta g^{\mu\nu}}$.
Varying the action (\ref{action2}) with respect to the metric tensor $g^{\mu\nu}$,
\begin{equation}\label{frt}
f_R(R,T)R_{\mu\nu} -\frac{1}{2}f(R,T)g_{\mu\nu} + (g_{\mu\nu}
\square
-\triangledown_\mu\triangledown_\nu)f_R(R,T)=8\pi T_{\mu\nu} -f_T(R, T)T_{\mu\nu}- f_T(R,T) \Theta_{\mu\nu},
\end{equation}
where $f_R(R,T) \equiv \dfrac{\partial f(R, T)}{\partial R}$ and
$ f_T(R,T) \equiv \dfrac{\partial f(R,T)}{\partial T}.$ Note that, if we take $f(R, T)=R$ and $f(R, T)=f(R)$, then the equations \eqref{frt} become the   Einstein field equations of general relativity and $f(R)$ gravity, respectively. In the present work, we assume that the stress-energy tensor is defined as
\begin{equation}\label{energy}
  T_{\mu\nu}=(p+\rho)u_\mu u_\nu -p g_{\mu\nu},
\end{equation}
and the matter Lagrangian can be taken as $\mathcal{L}=-p$. The
four velocity $u_{\mu}$ satisfies the conditions $u_{\mu}u^{\mu}=1$ and $u^{\mu}\triangledown_{\nu}u_{\mu}=0$. If the matter source is a perfect fluid, then $\Theta_{\mu\nu}=-2T_{\mu\nu}-pg_{\mu\nu}$.

In this  study, we shall consider $f(R, T)=f(R)+2f(T)$, where $f(R)$ is a function of $R$ and $f(T)$ is a function of the trace of the energy momentum tensor, i.e. $T=\rho-3p$.


\section{$f(R, T)=R+\lambda R^2+2\beta\ln(T)$ and field equations}
In this paper, the novel $f(R, T)$ function is defined as
\begin{equation}\label{newfunc}
f(R, T)=R+\lambda R^2+2\beta\ln(T),
\end{equation}
where $\lambda$, $\beta$ are constants and $T=\rho-3p>0$. The dimensions of $\lambda$ and $\beta$ are [Time]$^{2}$ and [Time]$^{-2}$, respectively. The units of $\lambda$ and $\beta$ are taken as sec$^{2}$ and sec$^{-2}$.

From the above choice of $f(R, T)$, we come to know that $\rho-3p>0$,  otherwise the function $f(R, T)$ would not be well defined. Therefore, $\rho-3p>0$ is mandatory. Subsequently, we can say that $\rho+3p>0$, provided $p>0$. Now, if we are able to
 show that $\rho >0$, then all energy conditions will be satisfied, namely the Null Energy Condition (NEC), the Weak Energy Condition (WEC), and the Strong Energy Condition (SEC). Consequently, we will be able to say that our model does not contain any exotic type of matter for this particular choice of the $f(R, T)$ function. The spacetime of the model is assumed to be the flat Fiedmann-Lemaitre-Robertson-Walker (FLRW) metric, which is defined as
\begin{equation}\label{metric}
  ds^2=dt^2-a^2(t)(dx^2+dy^2+dz^2).
\end{equation}
 Using Eqs. \eqref{energy}, \eqref{newfunc} and \eqref{metric} in the field equation \eqref{frt}, the explicit form of the field equations are obtained as
 \begin{equation}\label{e1}
 	3\left(\frac{\dot{a}}{a}\right)^2 -\frac{18\lambda}{a^4}[\ddot{a}^2a^2+2\dot{a}\dddot{a}a^2-5\dot{a}^4+2a\dot{a}^2\ddot{a}]=8\pi\rho+2\beta\frac{\rho+p}{\rho-3p}-\beta\ln(\rho-3p)
 \end{equation}
 and
 \begin{equation}\label{e2}
 \frac{2\ddot{a}}{a}+\left(\frac{\dot{a}}{a}\right)^2-\frac{6\lambda}{a^4}\Big[5\dot{a}^4-12a\dot{a}^2\ddot{a}+a^2\ddot{a}^2+4a^2\dot{a}\dddot{a}+2a^3\ddddot{a}\Big]=-8\pi p-\beta\ln(\rho-3p),
 \end{equation}
where the overhead dot denotes the derivative with respect to time `t'.

By taking $\lambda=0$, the field equations \eqref{e1} and \eqref{e2} reduce to
\begin{equation}\label{f1}
  3\left(\frac{\dot{a}}{a}\right)^2=8\pi \rho +\frac{2\beta (\rho+p)}{\rho-3p}-\beta \ln (\rho-3p),
\end{equation}
\begin{equation}\label{f2}
  \frac{2\ddot{a}}{a}+\left(\frac{\dot{a}}{a}\right)^2=-\beta\ln (\rho-3p)-8\pi p.
\end{equation}
%

The  covariant
divergence of \eqref{frt} implies
\begin{equation}\label{diverg}
\triangledown^{\mu}T_{\mu\nu}=\frac{f_{T}(R,T)}{8\pi-f_{T}(R,T)}\Bigg[(T_{\mu\nu}+\Theta_{\mu\nu})\triangledown^{\mu}\ln f_{T}(R,T) +\triangledown^{\mu}\Theta_{\mu\nu}-\frac{1}{2}g_{\mu\nu}\triangledown^{\mu}T\Bigg],
\end{equation}
\noindent
which is not equal to zero. It happens due to the presence of higher order derivatives of the energy-momentum tensor in the field equations. Hence the theory might be plagued by divergences at astrophysical scales. It looks to be a problem with some other higher order derivatives theories as well that contain higher order terms of energy-momentum tensor. Nevertheless, to deal this problem, one can put some constraints to Eq. \eqref{diverg} to obtain standard conservation equation\cite{Sharif1}.
\section{Model investigations}

Nowadays, the bouncing problem is one of the fascinating parts of the study of cosmological dynamics in modified gravity, because the big bang singularity could be avoided by a big bounce\cite{Molina, Cai}. The indication of the bouncing universe is: the size of the scale factor is contracted to a finite volume, not necessarily zero, and then blows up. Subsequently, it delivers a conceivable solution to the singularity problem of the standard big bang model.
To become a bounce, there must be some finite point of time at which the size of the universe attains minimum. Let us take this time to be $t=t_0$. As per the fundamental rule of differential calculus, a function $`f'$ attains minimum at say $t=t_0$, provided $f$ satisfies $\dot{f}(t_0)=0$ and $\ddot{f}(t_0)>0$. Hence, the behavior of the scale factor at $t_0$ must be $\dot{a}(t_0)=0$ and $\ddot{a}(t_0)>0$. Precisely we can say that, if $t_0$ is a bounce point, then the scale factor $a(t)$ decreases, i.e. $\dot{a}(t)<0$, for $t<t_0$ and the scale factor $a(t)$ increases, i.e. $\dot{a}(t)>0$, for $t>t_0$, locally. Equivalently in the bouncing model the hubble parameter $H$ runs across zero from $H<0$ to $H>0$ and $H=0$ at the bouncing point. Subsequently, for bouncing problem, we must required $a(t)>a(t_0)$ for $t\ne t_0$ locally. In addition to that, the violation of the Null Energy Condition (NEC) is required for a period of time inside the neighbourhood of the bounce point in general relativity within the frame-work of spatially flat 4-dimensional FLRW model. Furthermore, the EoS parameter $\omega$ of the matter content present in the universe must experience a phase switch from $\omega<-1$ to $\omega>-1$, to enter into the hot big bang age after the bounce\cite{Molina, Cai}. Note that bounces in $f(R)$ gravity were investigated in\cite{Odintsov:2015uca, Odintsov:2015zua, Odintsov:2015zza, Odintsov:2015ynk}.

 Cai et al. \cite{Cai} studied a bouncing universe filled with Quintom matter which is described by the EoS $w(t)=-r-\frac{s}{t^2}$, where $r$ and $s$ are the parameters with $r<1$ and $s>0$. Here, we are motivated by their work and defined EoS of two different types, which represent Quintom matter, to study the bouncing scenarios of the universe in modified gravity.
The intention of this section is to study the various cosmological dynamics under bouncing situations by defining two new equation of state (EoS) parameters and one new prametrized form of the Hubble parameter.

Let us start with a study on the possibility of obtaining the bouncing solution described by the following EoS:
\begin{equation}\label{eos1}
  \omega(t)=-\frac{k\ln(t+\epsilon)}{t}-1,
\end{equation}
where $\epsilon$ is very small and $k$ is any arbitrary constant. We see from Eq. \eqref{eos1} that $\omega$ varies from negative infinity at $t=0$ to the cosmological constant at $t=1-\epsilon$, and crosses this boundary, eventually it comes back to again cosmological constant for $t\to\infty$.

To study the bouncing solution of cosmological models, we define one more EoS as follows:
\begin{equation}\label{eos2}
  \omega(t)=\frac{r}{\ln t}-s,
\end{equation}
  where $r$ and $s$ are parameters, we require $r<0$ and $s>0$. We see from Eq. \eqref{eos2} that $\omega$ varies from negative at $t=0$ to the cosmological constant at $t=e^{\frac{r}{s-1}}$, and crosses this boundary, eventually it comes back to again cosmological constant for $t\to\infty$ and $s=1$.

Now, we would like to define the parametric form of the Hubble parameter $H$ that describes the expansion
of the universe and helps us to accomplish some remarkable bouncing solutions. We parametrize the
functional form of the Hubble parameter, which is defined as
\begin{equation}\label{h1}
  H(t)=\alpha\sin(\xi t)h(t),
\end{equation}
where $\xi$ and $\alpha$ are arbitrary constants, and $h(t)$ is any smooth function.
Looking at the novel proposed form of the Hubble parameter, we can observe that the trigonometric function $\sin(kt)$ vanishes at $t=0, \frac{\pi}{k}, \frac{2\pi}{k}, \cdots, \frac{n\pi}{k}$,  implying that the scale factor must take a constant value at these points and the second function $h(t)$ should not vanish at those points. Now, we will have to choose $h(t)$, which could be any algebraic, rational, exponential, transcendental, or periodic function, in such a way that it should be smooth and non-vanishing at those points.
 In our study, we consider a specific form of $h(t)$, which is defined as
\begin{equation}\label{h2}
  h(t)=e^{\zeta t},
\end{equation}
where $\zeta$ is any arbitrary constant.
Hence, the complete parametrize form of the Hubble parameter $H(t)$ is as follows:
\begin{equation}\label{h3}
  H(t)=\alpha\sin(\xi t)e^{\zeta t}.
\end{equation}

	\begin{figure}{}
	\centering
	\includegraphics[scale=.8]{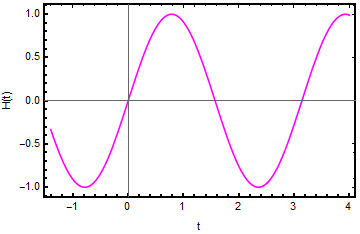}\hspace{.05cm}
	\caption{Hubble parameter $H(t)$ versus $t$: We have considered two bouncing points, one is $t=0$ and the second one is $t=3.2$. At the bouncing point $t=0$, the Hubble parameter attains zero. It has negative values for $t<0$ and positive values for $t>0$. The behavior of the Hubble parameter is the same near the second bounce point, as well. For this plot, $\xi=2$, $\zeta=-.0001$ and $\alpha=1$ are taken.}
	\end{figure}
	
	\begin{figure}
	\centering
\includegraphics[scale=.8]{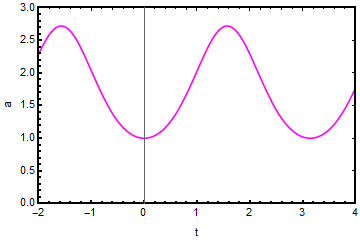}\hspace{.05cm}	
	\caption{Scale factor $a$ versus $t$: Here, we have considered two bouncing points, one is $t=0$ and the second one is $t=3.2$. At the bouncing point $t=0$, the scale factor $a$ attains a minimum and the scale factor $a(t)>a(0)$, for $t\ne 0$. Consequently, the scale factor decreases for $t<0$, i.e. $\dot{a}(t)<0$, for $t<0$ and increases for $t>0$, i.e. $\dot{a}(t)>0$, for $t>0$. The behavior of the scale factor is exactly the same near the second bounce point, as well. For this figure, $\xi=2$, $\zeta=-0.0001$ and $\alpha=1$ are taken.}
		\end{figure}

\begin{figure}{}
	\centering
\includegraphics[scale=.8]{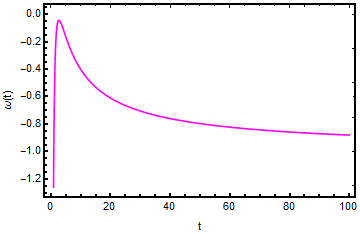}\hspace{.05cm}
	\caption{The equation of state parameter $\omega(t)=-\frac{k\ln(t+\epsilon)}{t}-1$ versus $t$: The value of $\omega$ tends to negative infinity at the bouncing point $t=0$, while after the bouncing point $t=0$, the value of $\omega$ varies from $-1$ to $0$, i.e. $-1<\omega<0$. However, $\omega\to\infty$, for $t\to\infty$, which indicates a late time cosmic acceleration, as the universe is dominated by a cosmological constant. For this plot, $k=-2.6$ and $\epsilon=-0.0004$ are used.}
	\end{figure}
	
	\begin{figure}{}
	\centering
\includegraphics[scale=.8]{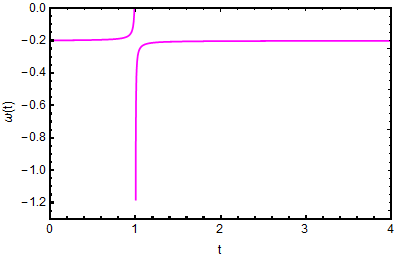}\hspace{.05cm}
	\caption{The equation of state parameter $\omega(t)=\frac{r}{\ln t}-s$ versus $t$: The value of the equation of state parameter $\omega$ lies between -1 and 0 within a neighbourhood of bouncing point $t = 0$. After that, the value of $\omega$ lies between -1.2 and 0. For this figure, $r=-0.0025$ and $s=0.2$ are taken. }
	\end{figure}
	
 \noindent
The evolution of the scale factor $a(t)$  comes out be
  \begin{equation}\label{a1}
    a=\kappa \exp\left(\alpha\frac{e^{\zeta t}[\zeta\sin(\xi t)-\xi\cos(\xi t)]}{\zeta^2+\xi^2}\right),
  \end{equation}
where $\kappa$ is an integration constant.

From Eqs. \eqref{h3} and \eqref{a1}, we observe that the Hubble parameter $H(t)$ vanishes at $t=0, \frac{\pi}{k}, \frac{2\pi}{k}, \ldots, \frac{n\pi}{k}$, so we can choose $t=0$ as one bouncing point.  One can see that our solution provides
a picture of evolution of the universe with a contracting phase for $t<0$, and then bouncing at $t=0$ to the expanding phase for $t>0$.

We would like to investigate the model defined in the preceding section for two different cases: \\

 \noindent
 \textbf{Case I:} $\omega(t)=-\frac{k\ln(t+\epsilon)}{t}-1$.\\

 \noindent
\textbf{Subcase I(a):} $\lambda=0$.\\
\begin{equation}
\rho=\frac{t}{4 \pi}\Bigg[-\frac{\beta}{3 k \ln (t+\epsilon )+4 t}+\frac{\alpha   e^{\zeta  t} (\zeta  \sin (\xi  t)+\xi  \cos (\xi  t))}{k \ln(t + \epsilon)}\Bigg].
\end{equation}

\begin{equation}
p=-\frac{(k \ln (t+\epsilon )+t)}{4\pi}\Bigg[ -\frac{\beta}{3 k \ln (t+\epsilon )+4 t}+\frac{\alpha   e^{\zeta  t} (\zeta  \sin (\xi  t)+\xi  \cos (\xi  t))}{k \ln(t + \epsilon)}\Bigg].
\end{equation}

\begin{figure}{}
	\centering
	\subfigure[Density ($\rho$): This plot indicates that the energy density is positive within a neighborhood of the bouncing point $t=0$. ]{\includegraphics[width=7cm,height=5cm]{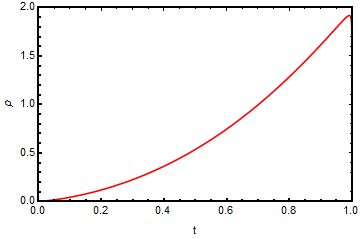}}\hspace{.5cm}
	\subfigure[Pressure ($p$): This plot indicates that the pressure is negative. ]{\includegraphics[width=7cm,height=5cm]{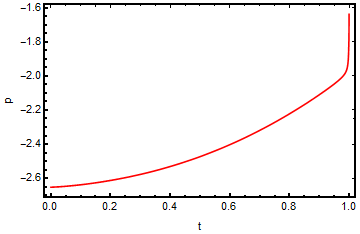}}
	\subfigure[NEC ($\rho+p$): It is observed that the term $\rho+p$ is positive near the bounce point $t=0$, which indicates that the NEC is satisfied.
] {\includegraphics[width=7cm,height=5cm]{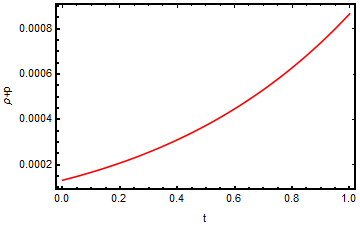}}\hspace{.5cm}
	\subfigure[SEC ($\rho+3p$): This figure shows that $\rho+3p>0$ near the throat. From figures  (c) and (d), the SEC is satisfied near the bounce point $t=0$. ]{\includegraphics[width=7cm,height=5cm]{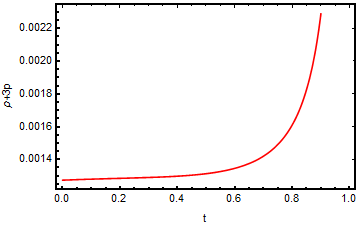}}
	\subfigure[DEC ($\rho-|p|$): For the DEC, we required $\rho>0$ and $\rho-|p|>$. Hence, from figures (a) and (e), it is observed that the DEC is satisfied within a neighborhood of the bouncing point $t=0$. ]{\includegraphics[width=7cm,height=5cm]{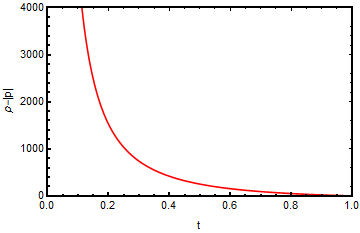}}\hspace{.5cm}
	\subfigure[Stress energy tensor ($T$): In this figure, it is shown that the term $\rho-3p>0$, which indicates that the newly defined $f(R, T)$ function is well defined. ]{\includegraphics[width=7cm,height=5cm]{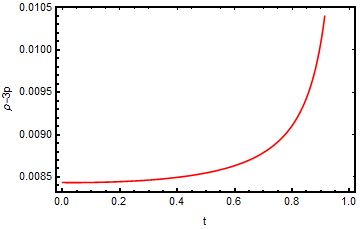}}
	\caption{Plots of density, pressure, NEC, SEC, DEC \& stress energy tensor for  $\omega(t)=-\frac{k\ln(t+\epsilon)}{t}-1$, with $\lambda=0$, $\xi=2$, $\zeta=-0.0001$, $\alpha=1$, $\beta=-100$, $k=-2.6$ and $\epsilon=-0.0004$}
\end{figure}

\noindent
\textbf{Subcase I(b):} $\lambda\neq 0$
\begin{eqnarray}
\rho&=&\frac{t}{4 \pi  k \ln (t+\epsilon ) (3 k \ln (t+\epsilon )+4 t)}\Bigg[48 \alpha^3 \lambda  e^{3 \zeta  t} \sin ^2(\xi  t) (3 k \ln (t+\epsilon )+4 t) (\zeta  \sin (\xi  t)\nonumber\\
&+&\xi  \cos (\xi  t))+3 \alpha ^2 \lambda  e^{2 \zeta  t} (3 k \ln (t+\epsilon )+4 t) \left(-5 \zeta ^2+\xi ^2-10 \zeta  \xi  \sin (2 \xi  t)+5 (\zeta -\xi )\right. \nonumber\\
&\times&\left.(\zeta +\xi ) \cos (2 \xi  t)\right)+\alpha  e^{\zeta  t} (3 k \ln (t+\epsilon )+4 t) \left(\zeta  \left(-6 \zeta ^2 \lambda +18 \lambda  \xi ^2+1\right) \sin (\xi  t)\right.\nonumber\\
&+&\left.\xi  \left(6 \lambda  \left(\xi ^2-3 \zeta ^2\right)+1\right) \cos (\xi  t)\right)-\beta  k \ln (t+\epsilon )\Bigg],
\end{eqnarray}

\begin{eqnarray}
p&=&-\frac{(k \ln (t+\epsilon )+t)}{4 \pi  k \ln (t+\epsilon ) (3 k \ln (t+\epsilon )+4 t)}\Bigg[48 \alpha^3 \lambda  e^{3 \zeta  t} \sin ^2(\xi  t) (3 k \ln (t+\epsilon )+4 t) (\zeta  \sin (\xi  t)\nonumber\\
&+&\xi  \cos (\xi  t))+3 \alpha ^2 \lambda  e^{2 \zeta  t} (3 k \ln (t+\epsilon )+4 t) \left(-5 \zeta ^2+\xi ^2-10 \zeta  \xi  \sin (2 \xi  t)+5 (\zeta -\xi )\right. \nonumber\\
&\times&\left.(\zeta +\xi ) \cos (2 \xi  t)\right)+\alpha  e^{\zeta  t} (3 k \ln (t+\epsilon )+4 t) \left(\zeta  \left(-6 \zeta ^2 \lambda +18 \lambda  \xi ^2+1\right) \sin (\xi  t)\right.\nonumber\\
&+&\left.\xi  \left(6 \lambda  \left(\xi ^2-3 \zeta ^2\right)+1\right) \cos (\xi  t)\right)-\beta  k \ln (t+\epsilon )\Bigg].
\end{eqnarray}

\begin{figure}{}
	\centering
	\subfigure[Density ($\rho$): In this plot, it is observed that the energy density in positive around the bouncing point $t=0$.]{\includegraphics[width=7cm,height=5cm]{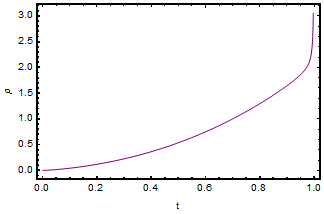}}\hspace{.5cm}
	\subfigure[Pressure ($p$): In this plot, it is observed that the pressure is negative, which could indicate the cause of the accelerated expansion of the universe.]{\includegraphics[width=7cm,height=5cm]{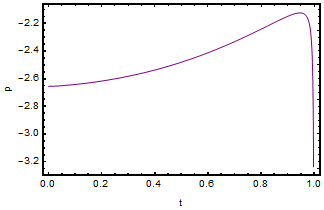}}
	\subfigure[NEC ($\rho+p$): In this plot, it is observed that the NEC is satisfied. Moreover, including figures (a) and (c), we can say that the WEC is satisfied as well as near the bouncing point $t=0$. ]{\includegraphics[width=7cm,height=5cm]{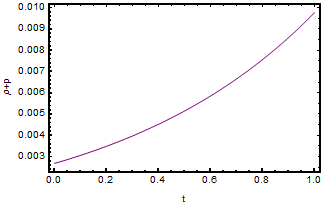}}\hspace{.5cm}
	\subfigure[SEC ($\rho+3p$): From  figures (c) and (d), we can say that the SEC is satisfied within a neighborhood of the bouncing point $t=0$.]{\includegraphics[width=7cm,height=5cm]{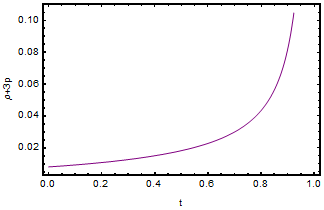}}
	\subfigure[DEC ($\rho-|p|$): In figures (a) and (d), we can say that the DEC is satisfied within a neighborhood of the bouncing point $t=0$.]{\includegraphics[width=7cm,height=5cm]{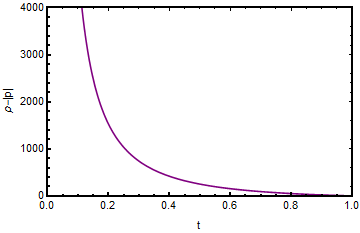}}\hspace{.5cm}
	\subfigure[Stress energy tensor ($T$): In this plot, it is observed that  $T=\rho-3p$ is positive, which means that the newly defined $f(R, T)$ function is well defined.]{\includegraphics[width=7cm,height=5cm]{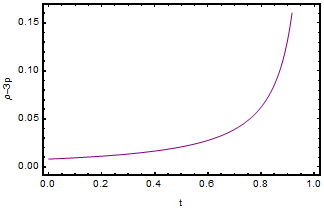}}
	\caption{ Plots of density, pressure, NEC, SEC, DEC and stress energy tensor for $\omega(t)=-\frac{k\ln(t+\epsilon)}{t}-1$, with $\lambda=1$, $\xi=2$, $\zeta=-0.0001$, $\alpha=1$, $\beta=-100$, $k=-2.6$ and $\epsilon=-0.0004$}
		\end{figure}

\noindent
\textbf{Case II:} $\omega(t)=\frac{r}{\ln t}-s$.\\

\noindent
\textbf{Subcase II(a):} $\lambda=0$.\\
\begin{eqnarray}
\rho&=&\frac{\ln (t)}{4 \pi } \left(-\frac{\alpha  e^{\zeta  t} (\zeta  \sin (\xi  t)+\xi  \cos (\xi  t))}{r-s \ln (t)+\ln (t)}-\frac{\beta}{-3 r+3 s \ln (t)+\ln (t)}\right),
\end{eqnarray}

\begin{eqnarray}
p&=&\frac{\ln (t)}{4 \pi }\Bigg(\frac{r}{\ln t}-s\Bigg)\left(-\frac{\alpha  e^{\zeta  t} (\zeta  \sin (\xi  t)+\xi  \cos (\xi  t))}{r-s \ln (t)+\ln (t)}-\frac{\beta}{-3 r+3 s \ln (t)+\ln (t)}\right).
\end{eqnarray}
\begin{figure}{}
	\centering
	\subfigure[Density ($\rho$): In this figure, it is shown that the density $\rho$ is positive within a neighborhood of the bouncing point $t=0$.]{\includegraphics[width=7cm,height=5cm]{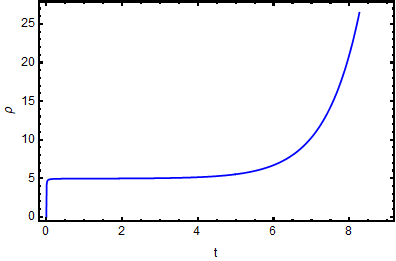}}\hspace{.5cm}
	\subfigure[Pressure ($p$): In this figure, it is observed that the pressure is negative throughout the evolution of the universe. ]{\includegraphics[width=7cm,height=5cm]{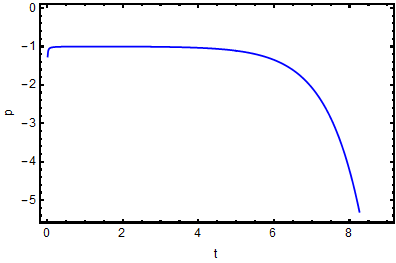}}
	\subfigure[NEC ($\rho+p$): In this figure, it is observed that the NEC is satisfied, subsequently, from figures (a) and (c), we can say that the WEC condition is satisfied, as well.]{\includegraphics[width=7cm,height=5cm]{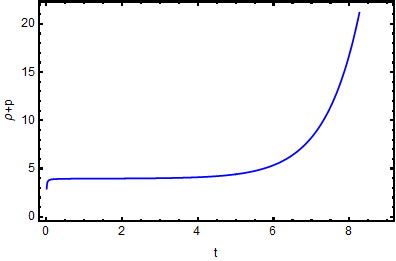}}\hspace{.5cm}
	\subfigure[SEC ($\rho+3p$): We can observe from the figures (c) and (d) that both $\rho+p$ and $\rho+3p$ are positive, which provides a validation of the SEC within a neighborhood of the bouncing point $t=0$. ]{\includegraphics[width=7cm,height=5cm]{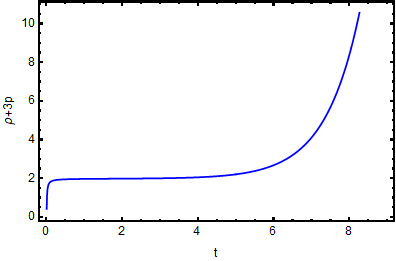}}
	\subfigure[DEC ($\rho-|p|$): From figures (a) and (e), it is observed that the DEC is satisfied within a neighborhood of the bouncing point $t=0$.]{\includegraphics[width=7cm,height=5cm]{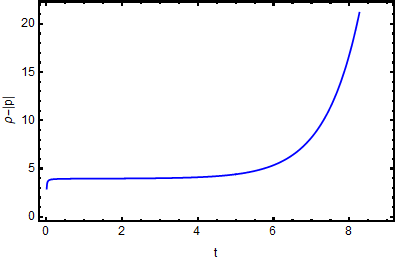}}\hspace{.5cm}
	\subfigure[Stress energy tensor ($T$): This figure indicates that $T=\rho-3p$ is positive throughout the evolution of the universe, which accounts for the validity of the $f(R, T)$ function. ]{\includegraphics[width=7cm,height=5cm]{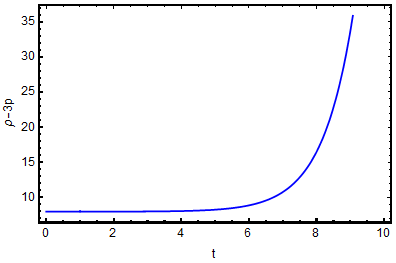}}
	\caption{Plots of density, pressure, NEC, SEC, DEC \& stress energy tensor  and $\omega(t)=\frac{r}{\ln t}-s$, with $\lambda=0$, $\xi=2$, $\zeta=-0.0001$, $\alpha=1$, $\beta=-100$, $r=-0.001$ and $s=0.2$. }
		\end{figure}

\textbf{Subcase II(b):} $\lambda\neq0$\\
\begin{eqnarray}
\rho&=&-\frac{\log (t)}{4 \pi  (r-s \log (t)+\log (t)) (3 r-(3 s+1) \log (t))} \Bigg[-\beta  r+48 \alpha ^3 \zeta  \lambda  e^{3 \zeta  t} \sin ^3(\xi  t)\nonumber\\
&\times&(3 r-(3 s+1) \log (t))+\alpha  e^{\zeta  t} \sin (\xi  t) (3 r-(3 s+1) \log (t)) \left(-6 \zeta ^3 \lambda +18 \zeta  \lambda  \xi ^2+\zeta \right.\nonumber\\
&+&24 \left.\alpha ^2 \lambda  \xi  e^{2 \zeta  t} \sin (2 \xi  t)\right)-6 \alpha ^2 \lambda  \left(5 \zeta ^2-3 \xi ^2\right) e^{2 \zeta  t} \sin ^2(\xi  t) (3 r-(3 s+1) \log (t))\nonumber\\
&-&12 \alpha ^2 \lambda  \xi ^2 e^{2 \zeta  t} \cos ^2(\xi  t) (3 r-(3 s+1) \log (t))+\alpha  \xi  e^{\zeta  t} \left(-18 \zeta ^2 \lambda +6 \lambda  \xi ^2+1\right) \cos (\xi  t)\nonumber\\
&\times& (3 r-(3 s+1) \log (t))-90 \alpha ^2 \zeta  \lambda  \xi  r e^{2 \zeta  t} \sin (2 \xi  t)+90 \alpha ^2 \zeta  \lambda  \xi  s e^{2 \zeta  t} \log (t) \sin (2 \xi  t)\nonumber\\
&+&\beta  s \log (t)+30 \alpha ^2 \zeta  \lambda  \xi  e^{2 \zeta  t} \log (t) \sin (2 \xi  t)-\beta  \log (t)\Big],
\end{eqnarray}

\begin{eqnarray}
p&=&-\frac{\log (t)\left(\frac{r}{\ln t}-s\right)}{4 \pi  (r-s \log (t)+\log (t)) (3 r-(3 s+1) \log (t))} \Bigg[-\beta  r+48 \alpha ^3 \zeta  \lambda  e^{3 \zeta  t} \sin ^3(\xi  t)\nonumber\\
&\times&(3 r-(3 s+1) \log (t))+\alpha  e^{\zeta  t} \sin (\xi  t) (3 r-(3 s+1) \log (t)) \left(-6 \zeta ^3 \lambda +18 \zeta  \lambda  \xi ^2+\zeta \right.\nonumber\\
&+&24 \left.\alpha ^2 \lambda  \xi  e^{2 \zeta  t} \sin (2 \xi  t)\right)-6 \alpha ^2 \lambda  \left(5 \zeta ^2-3 \xi ^2\right) e^{2 \zeta  t} \sin ^2(\xi  t) (3 r-(3 s+1) \log (t))\nonumber\\
&-&12 \alpha ^2 \lambda  \xi ^2 e^{2 \zeta  t} \cos ^2(\xi  t) (3 r-(3 s+1) \log (t))+\alpha  \xi  e^{\zeta  t} \left(-18 \zeta ^2 \lambda +6 \lambda  \xi ^2+1\right) \cos (\xi  t)\nonumber\\
&\times& (3 r-(3 s+1) \log (t))-90 \alpha ^2 \zeta  \lambda  \xi  r e^{2 \zeta  t} \sin (2 \xi  t)+90 \alpha ^2 \zeta  \lambda  \xi  s e^{2 \zeta  t} \log (t) \sin (2 \xi  t)\nonumber\\
&+&\beta  s \log (t)+30 \alpha ^2 \zeta  \lambda  \xi  e^{2 \zeta  t} \log (t) \sin (2 \xi  t)-\beta  \log (t)\Big].
\end{eqnarray}

\begin{figure}{}
	\centering
	\subfigure[Density ($\rho$): The energy density is seen to be positive, within a neighbourhood of the bouncing point $t=0.$ ]{\includegraphics[width=7cm,height=5cm]{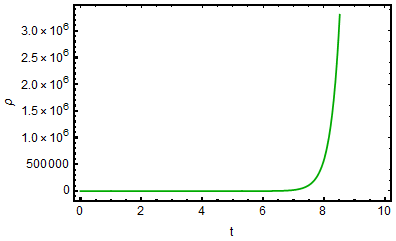}}\hspace{.5cm}
	\subfigure[Pressure ($p$): The pressure is negative within a neighbourhood of the bouncing point $t=0$. ]{\includegraphics[width=7cm,height=5cm]{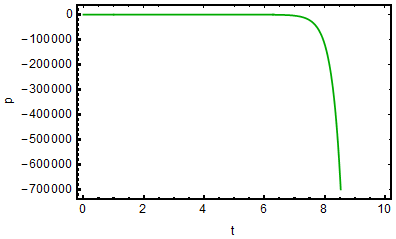}}
	\subfigure[NEC ($\rho+p$): It is shown that $\rho+p>0$, which indicates that the NEC is satisfied near the bouncing point. From  Fig. (a), we can see that the energy density $\rho$ is positive, hence the WEC is satisfied, as well, within a neighbourhood of the bouncing point.
 ]{\includegraphics[width=7cm,height=5cm]{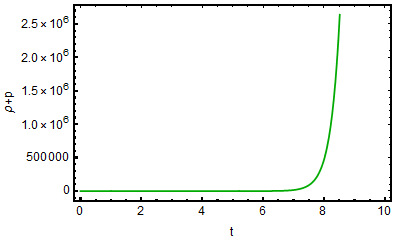}}\hspace{.5cm}
	\subfigure[SEC ($\rho+3p$): This plot indicates that $\rho+3p>0$ near the bounce point $t=0$. From the figures (c) and (d), we can say that $\rho+p>0$ and $\rho+3p>0$, which indicates that the SEC is satisfied within a neighbourhood of the bouncing point.]{\includegraphics[width=7cm,height=5cm]{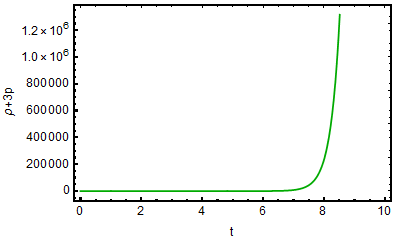}}
	\subfigure[DEC ($\rho-|p|$): From figures (a) and (e), we confirm that the DEC is satisfied within a neighbourhood of the bouncing point $t=0$.]{\includegraphics[width=7cm,height=5cm]{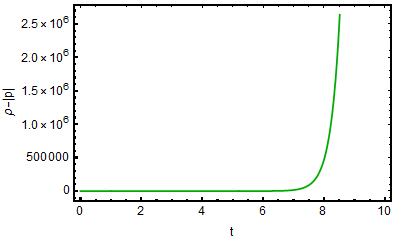}}\hspace{.5cm}
	\subfigure[Stress energy tensor ($T$): In this plot, it is observed that $T=\rho-3p>0$, which indicates that our newly defined $f(R, T)$ function is well defined.]{\includegraphics[width=7cm,height=5cm]{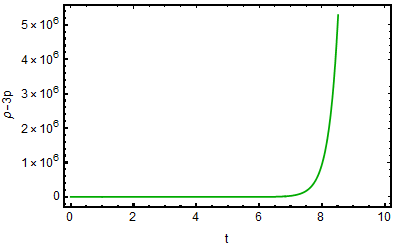}}
	\caption{Plots of density, pressure, NEC, SEC, DEC \& stress energy tensor for $\omega(t)=\frac{r}{\ln t}-s$, with $\lambda=0$, $\xi=2$, $\zeta=-0.0001$, $\alpha=1$, $\beta=-100$, $r=-0.001$ and $s=0.2$.}
		\end{figure}
		
\section{Hubble parameter versus redshift}		
In this section, a differential equation for the Hubble parameter with respect to the redshift is obtained. The values of the parameters  present in the differential equation are estimated using the $\chi^2$ test. Then, the differential equation is solved and  theoretical values for the Hubble parameter are calculated. Further, the theoretical and  observational values of the Hubble parameter are compared.\\

Differentiating \eqref{h3} with respect to $t$,
\begin{equation}\label{1}
\frac{dH}{dt}=\alpha e^{\zeta t}(\xi\cos \xi t+\zeta\sin \xi t).
\end{equation}
From Eq. \eqref{1} and using the relation $1+z=\frac{a_0}{a}$, we have obtained
\begin{equation}\label{2}
(1+z)H(z)\frac{dH(z)}{dz}+\zeta H(z)=-\alpha\xi e^{\zeta t}\cos \xi t.
\end{equation}
Now, using Eqns. \eqref{h3} and \eqref{a1}, we get
\begin{equation}\label{H}
(1+z)H(z)\frac{dH(z)}{dz}+2\zeta H(z)+(\xi^2+\zeta^2)\ln(1+z)=0.
\end{equation}

The differential equation \eqref{H} contains two parameters, $\zeta$ and $\xi$. The values of these parameters are obtained by minimizing
\begin{equation}
\chi^2(\zeta,\xi)=\sum_{i=1}^{29}\frac{[H_{th}(\zeta,\xi;z_i)-H_{obs}(z_i)]^2}{\sigma^2_{H(z_i)}},
\end{equation}
where $H_{th}$ and $H_{obs}$ denote theoretical and observed values of Hubble parameter, respectively and $\sigma_{H(z_i)}$ denotes the standard error for each observed value. The present value of the Hubble parameter is considered as $H_0=67.8$ Km/s/Mpc \cite{Planck}. A total of 29 observational Hubble parameter data  are taken from \cite{Jimenez, Simon, Stern, Moresco, Blake, Zhang, Moresco1, Delubac}. On the $\zeta-\xi$ plane, the likelihood contours for $1\sigma$, $2\sigma$ and $3\sigma$ errors are drawn in Fig. (9).
The values of the parameters $\zeta$ and $\xi$ are found to be equal to -46 and 1.6 respectively for a minimum value 123.567 of $\chi^2$.
For  $\zeta=-46$ and $\xi=1.6$, a numerical solution for the  non-linear differential equation \eqref{H} is obtained and plotted in Fig. (10) with respect to the redshift through a continuous curve.
  In this figure, the bars correspond to the error between theoretical and observational values of the Hubble parameter. Clearly, it provides a good fit to the experimental Hubble parameter data with $\zeta=-46$ and $\xi=1.6$.  Thus, it enhances the importance of the newly defined Hubble parameter for the present model.
\begin{figure}{}
	\centering
\includegraphics[scale=.8]{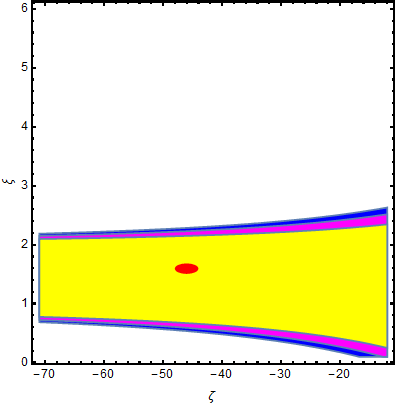}\hspace{.05cm}
	\caption{1$\sigma$ (yellow shaded), 2$\sigma$ (mergenta shaded) and 3$\sigma$ (blue shaded) likelihood contours  on the $\zeta-\xi$ plane.}
	\end{figure}
	
	\begin{figure}{}
	\centering
	\includegraphics[scale=.9]{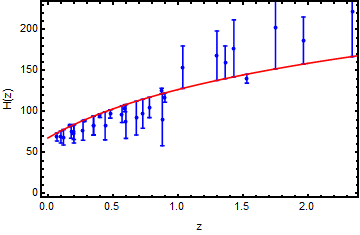}\hspace{.05cm}
	\caption{Hubble parameter $H(z)$ versus $z$: We have considered the present value of the Hubble parameter to be equal to 67.8 Km/s/Mpc, besides 29 observed Hubble data, $\zeta=-46$ and $\xi=1.6$ .}
	\end{figure}

\section{Results and discussion}
This work has been aimed at studying the cosmological dynamics  under a bouncing scenario within the framework of a spatially flat 4-dimensional FLRW model in a $f(R,T)$ theory of gravity. The $f(R,T)$ function is  defined as  $f(R,T)=R+\lambda R^2 +2\beta\ln(T)$, where $\lambda$ and $\beta$ are dimension-full constants and $T=\rho-3p>0$.  A parametric form of the Hubble parameter is also proposed as $H(t)=\alpha\sin(\xi t)e^{\zeta t}$, where $\alpha$, $\xi$ and $\zeta$ are arbitrary constants. Subsequently,  the scale factor has been obtained as $a(t)=\kappa\exp\Bigg(\alpha\frac{e^{\zeta t}[\zeta\sin(\xi t)-\xi\cos(\xi t)]}{\zeta^2+\xi^2}\Bigg)$, where $\kappa$ is an integration constant. Since $H(t)$ vanishes at $t=0, \frac{\pi}{k}, \frac{2\pi}{k},\ldots, \frac{n\pi}{k}$, the point $t=0$ is chosen as a first bouncing point. The Hubble parameter is plotted in Fig.(1). In the neighborhood of the bouncing point, we can observe a contraction phase for $t<0$, the bounce at $t=0$, and an expansion phase for $t>0$. The scale factor is normalized as  $a=1$ at the bouncing point. In Fig. (2), in the neighborhood of the bouncing point the scale factor is shown to decrease, for $t<0$, and  to increase, for $t>0$. Thus, the scale factor gets its non-zero and minimum value at $t=0$. \\

\noindent
Moreover, two forms of the EoS parameter are defined. The first form is $\omega(t)=-\frac{k\ln(t+\epsilon)}{t}-1$, where $\epsilon$ is very small and $k$ is an arbitrary  constant. It is plotted in Fig. (3) with respect to $t$. As $t$ increases from 0 to $1-\epsilon$, $\omega(t)$ increases from $-\infty$ to $-1$. Further, as $t$ increases from $1-\epsilon$ to 2.72, $\omega(t)$ increases towards -0.0436542, after what it decreases and approaches the value -1 as $t$ tends to infinity, which indicates that the universe is dominated by a cosmological constant at late time. Consequently, the late time accelerated expansion stage of our universe can be naturally realized in this model.  The second form of $\omega(t)$ is defined as  $\omega(t)=\frac{r}{\ln t}-s$, where $r$ and $s$ are constants, with  $r<0$ and $s>0$. It is plotted in Fig. (4) with respect to $t$, where  $\omega(t)$ is shown to vary from a negative value at $t=0$ to the cosmological constant at $t=1.00313$.
\\

\noindent
Using the background of $f(R,T)$ gravity, the Einstein's field equations are derived and, using the EoS parameters discussed above, these equations have been solved. The energy density $\rho$, pressure $p$, and the stress energy tensor $T=\rho-3p$ have been calculated. In addition to these, various combinations of $\rho$ and $p$, used in the following energy conditions, have been determined:\\
(i) Null energy condition (NEC): NEC $\Leftrightarrow \rho+p\geq 0$.\\
(ii) Weak energy condition (WEC): WEC $\Leftrightarrow \rho\geq 0, \rho+p\geq 0$.\\
(iii) Strong energy condition (SEC): SEC $\Leftrightarrow \rho+p\geq 0, \rho+3p\geq 0$.\\
(iv) Dominant energy condition (DEC): DEC $\Leftrightarrow \rho\geq 0, \rho-|p|\geq 0$.\\

\noindent
For each EoS parameter, the computation is performed in two subcases (a) $\lambda=0$ and (b) $\lambda\neq 0$. Summarizing, we have two cases: Case I: $\omega(t)=-\frac{k\ln(t+\epsilon)}{t}-1$, with subcases: I(a) $\lambda=0$ \& I(b) $\lambda\ne 0$, and Case II: $\omega(t)=\frac{r}{\ln t}-s$, with subcases: II(a) $\lambda=0$ \& II(b) $\lambda\ne 0$. In Subcases I(a) and I(b), the terms $\rho$, $p$, $\rho+p$, $\rho+3p$, $\rho-|p|$ and $\rho-3p$ are plotted for the variation of $t$ in a neighbourhood of the bouncing point $t=0$, i. e. from 0 to $1-\epsilon$ in Figs. (5) \& (6) respectively. Similarly, in Subcases II(a) and II(b), the results are plotted for the variation of $t$ from 0 to 10, in Figs. (7) \& (8), respectively. In each subcase, $\rho$, $\rho+p$, $\rho+3p$, $\rho-|p|$ and $\rho-3p$ are found to be positive. The positivity of $\rho-3p$ is necessary for the $f(R,T)$ function to be well defined and the positivity of $\rho$, $\rho+p$, $\rho+3p$ and $\rho-|p|$ implies  the absence of exotic matter in a neighborhood of the bouncing point, in our model, i.e. it only admits normal matter, satisfying the energy conditions in a neighborhood of the bouncing point.

Modern measurements of redshift and luminosity distance relations of type Ia supernovae indicate that the expansion of our universe is accelerating. This appears to be in strong disagreement with the standard picture of a matter dominated universe.  These observations can be accommodated theoretically by postulating that some form of an exotic matter with negative pressure dominates the present epoch of our universe. This exotic matter has been called Quintessence and behaves like a vacuum field energy with repulsive (anti-gravitational) character arising from the negative pressure.
In this paper, for each EoS, the pressure $p$ is found to be negative, i.e. the universe possesses a negative pressure, which could cause the accelerated expansion of the universe \cite{Kamenshchik}.
In short, by defining appropriate forms of the EoS parameter, the Hubble parameter and the $f(R,T)$ function, we have found the existence of a bouncing universe free from exotic matter and having a  non-vanishing and minimum scale factor at the bouncing point.
\\

\section{Conclusions}

In this paper, a non-singular bounce in a spatially flat 4-dimensional FLRW model has been explored by working with two EoS parameters and one novel parametric form of the Hubble parameter. Einstein's field equations have been obtained in the framework of $f(R,T)$ gravity with a newly proposed $f(R,T)=R+\lambda R^2 +2\beta\ln(T)$ function, where the condition $T=\rho-3p>0$ is required, so that the function $f(R, T)$ is well defined. From Figs. 5(f), 6(f), 7(f), and 8(f), it is observed that the stress energy momentum tensor $T=\rho-3p$ is positive. Hence, it turns out that our new function is well defined and its consistency is justified.

Usually, for general relativity, within the framework of the spatially flat 4-dimensional FLRW model, violation of the Null Energy Condition (NEC) is unavoidable for a period of time inside a neighbourhood of the bounce point \cite{Molina, Cai}. However, in the present study, we have obtained that all the energy conditions are satisfied within a neighborhood of the bouncing point, $t=0$. Therefore, we reach here  the interesting conclusion that the violation of the NEC is not always a necessary condition in modified gravity theories; at the very least, for the particular form of theory here considered, $f(R,T)=R+\lambda R^2 +2\beta\ln(T)$, it can be avoided.

Moreover, it is interesting to note that  pressure is negative within the neighborhood of the bouncing point
and, subsequently, it decreases to minus infinity throughout the evolution, which indicates that the universe may evolve indeed to a huge negative pressure stage. This negative pressure helps us to naturally realize the late time accelerated expansion of our universe. In our model, the best fit to the  experimental results for the Hubble parameter for different redshifts is determined for these values of  the parameters: $\zeta=-46$ and $\xi=1.6$.


\medskip

\noindent {\bf Acknowledgements}. This work was partially supported by MINECO (Spain), FIS2016-76363-P,  by the CPAN Consolider Ingenio 2010 Project, and by AGAUR (Catalan Government), project 2017-SGR-247. The work of author GCS is supported by CSIR Grant No.25(0260)/17/EMR-II. The authors are thankful to the anonymous reviewer for very constructive comments that helped to improve the quality of our work.

\newpage

\begin{table}[!h]
	\centering
	\caption{Hubble Parameter Observational data}
	\vspace{.5cm}
	\begin{tabular}{|c|c|c|c|c|}
		\hline
		S.No.& $z$ & $H(z)$ & $\sigma_i$&Reference\\
				\hline\hline
		1 & .090 & 69 & 12 & \cite{Jimenez}\\\hline
		2 & .17 & 83 & 8 & \cite{Simon}\\\hline		
		3 & .27 & 77 & 14 & \cite{Simon}\\\hline	
		4 & .4 & 95 & 17 & \cite{Simon}\\\hline	
		5 & .9 & 117 & 23 & \cite{Simon}\\\hline
	    6 & 1.3 & 168 & 17 & \cite{Simon}\\\hline
	    7 & 1.43 & 177 & 18 & \cite{Simon}\\\hline
	    8 & 1.53 & 140 & 14 & \cite{Simon}\\\hline
	    9 & 1.75 & 202 & 40 & \cite{Simon}\\\hline
	    10 & .48 & 97 & 62 & \cite{Stern}\\\hline
	    11 & .88 & 90 & 40 & \cite{Stern}\\\hline
	    12 & .179 & 75 & 4 & \cite{Moresco}\\\hline
	    13 & .199 & 75 & 5 & \cite{Moresco}\\\hline
	    14 & .352 & 83 & 14 & \cite{Moresco}\\\hline
	    15 & .593 & 104 & 13 & \cite{Moresco}\\\hline
	    16 & .68 & 92 & 8 & \cite{Moresco}\\\hline
	    17 & .781 & 105 & 12 & \cite{Moresco}\\\hline
	    18 & .875 & 125 & 17 & \cite{Moresco}\\\hline
	    19 & 1.037 & 154 & 20 & \cite{Moresco}\\\hline
	    20 & .44 & 82.6 & 7.8 & \cite{Blake}\\\hline
	    21 & .60 & 87.9 & 6.1 & \cite{Blake}\\\hline
	    22 & .73 & 97.3 & 7 & \cite{Blake}\\\hline
	    23 & .07 & 69 & 19.6 & \cite{Zhang}\\\hline
	    24 & .12 & 68.6 & 26.2 & \cite{Zhang}\\\hline
	    25 & .2 & 72.9 & 29.6 & \cite{Zhang}\\\hline
	    26 & .28 & 88.8 & 36.6 & \cite{Zhang}\\\hline
	    27 & 1.363 & 160 & 33.6 & \cite{Moresco1}\\\hline
	    28 & 1.965 & 186.5 & 50.4 & \cite{Moresco1}\\\hline
	    29 & 2.34 & 222 & 7 & \cite{Delubac}\\\hline
	\end{tabular}
\end{table}

\noindent
\textbf{{\Large Appendix}}\\

\noindent
\textbf{Case I:} $\omega(t)=-\frac{k\ln(t+\epsilon)}{t}-1$.\\

\noindent
\textbf{Subcase I(a):} $\lambda=0$.\\

\begin{equation}
\rho+p=-\frac{-\frac{k \beta \ln (t+\epsilon )}{3 k \ln (t+\epsilon )+4 t}+\alpha  e^{\zeta  t} (\zeta  \sin (\xi  t)+\xi  \cos (\xi  t))}{4 \pi },
\end{equation}

\begin{equation}
\rho+3p=-\frac{(3 k \ln (t+\epsilon )+2 t) \left(k\beta \ln (t+\epsilon )-\alpha  e^{\zeta  t} (3 k \ln (t+\epsilon )+4 t) (\zeta  \sin (\xi  t)+\xi  \cos (\xi  t))\right)}{4 \pi k \ln (t+\epsilon )(3 k \ln (t+\epsilon )+4 t)},
\end{equation}

\begin{eqnarray}
\rho-|p|&=&\frac{t}{4 \pi}\Bigg[-\frac{\beta}{3 k \ln (t+\epsilon )+4 t}+\frac{\alpha   e^{\zeta  t} (\zeta  \sin (\xi  t)+\xi  \cos (\xi  t))}{k \ln(t + \epsilon)}\Bigg]\\\nonumber
&-&\Bigg|\frac{(k \ln (t+\epsilon )+t) }{4\pi}\Bigg[\frac{\alpha  e^{\zeta  t} (\zeta  \sin (\xi  t)+\xi  \cos (\xi  t))}{k \ln (t+\epsilon )}-\frac{\beta}{3 k \ln (t+\epsilon )+4 t}\Bigg]\Bigg|,
\end{eqnarray}

\begin{eqnarray}
\rho-3p&=&\frac{1}{4 \pi k \log (t+\epsilon )}\Bigg[\alpha  e^{\zeta  t} (3 k \log (t+\epsilon )+4 t) (\zeta  \sin (\xi  t)+\xi  \cos (\xi  t))-k \beta \log (t+\epsilon )\Bigg].
\end{eqnarray}

\noindent
\textbf{Subcase I(b):} $\lambda\neq 0$.
\begin{eqnarray}
\rho+p&=&-\frac{1}{4 \pi }\Bigg[-\frac{\beta  k \ln (t+\epsilon )}{3k \ln (t+\epsilon )+4 t}+48 \alpha ^3 \lambda  e^{3 \zeta  t} \sin ^2(\xi  t) (\zeta  \sin (\xi  t)+\xi  \cos (\xi  t))+3 \alpha ^2 \lambda  e^{2 \zeta  t}\nonumber\\
&\times& \left(-5 \zeta ^2+\xi ^2-10 \zeta  \xi  \sin (2 \xi  t)+5 (\zeta -\xi ) (\zeta +\xi ) \cos (2 \xi  t)\right)+\alpha  e^{\zeta  t} \left(\zeta  \left(-6 \zeta ^2 \lambda\right.\right. \nonumber\\
&+&\left.\left.18 \lambda  \xi ^2+1\right) \sin (\xi  t)+\xi  \left(6 \lambda  \left(\xi ^2-3 \zeta ^2\right)+1\right) \cos (\xi  t)\right)\Bigg],
\end{eqnarray}

\begin{eqnarray}
\rho+3p&=&\frac{1}{4 \pi  k \ln (t+\epsilon ) (3 k \ln (t+\epsilon )+4 t)}\Bigg[(-3 k \ln (t+\epsilon )-2 t) \left(48 \alpha ^3 \lambda  e^{3 \zeta  t} \sin ^2(\xi  t)\right.\nonumber\\
&\times&\left. (3 k \ln (t+\epsilon )+4 t) (\zeta  \sin (\xi  t)+\xi  \cos (\xi  t))+3 \alpha ^2 \lambda  e^{2 \zeta  t} (3 k \ln (t+\epsilon )+4 t) \left(-5 \zeta ^2\right.\right.\nonumber\\
&+&\left.\left.\xi ^2-10 \zeta  \xi  \sin (2 \xi  t)+5 (\zeta -\xi ) (\zeta +\xi ) \cos (2 \xi  t)\right)+\alpha  e^{\zeta  t} (3 k \ln (t+\epsilon )+4 t)\right. \nonumber\\
&\times&\left(\zeta  \left(-6 \zeta ^2 \lambda +18 \lambda  \xi ^2+1\right) \sin (\xi  t)+\xi  \left(6 \lambda  \left(\xi ^2-3 \zeta ^2\right)+1\right) \cos (\xi  t)\right)\nonumber\\
&-&\left.\beta k \ln (t+\epsilon )\right)\Bigg],
\end{eqnarray}

\begin{eqnarray}
\rho-|p|&=&\frac{t}{4 \pi  k \ln (t+\epsilon ) (3 k \ln (t+\epsilon )+4 t)}\Bigg[48 \alpha^3 \lambda  e^{3 \zeta  t} \sin ^2(\xi  t) (3 k \ln (t+\epsilon )+4 t) (\zeta  \sin (\xi  t)\nonumber\\
&+&\xi  \cos (\xi  t))+3 \alpha ^2 \lambda  e^{2 \zeta  t} (3 k \ln (t+\epsilon )+4 t) \left(-5 \zeta ^2+\xi ^2-10 \zeta  \xi  \sin (2 \xi  t)+5 (\zeta -\xi )\right. \nonumber\\
&\times&\left.(\zeta +\xi ) \cos (2 \xi  t)\right)+\alpha  e^{\zeta  t} (3 k \ln (t+\epsilon )+4 t) \left(\zeta  \left(-6 \zeta ^2 \lambda +18 \lambda  \xi ^2+1\right) \sin (\xi  t)\right.\nonumber\\
&+&\left.\xi  \left(6 \lambda  \left(\xi ^2-3 \zeta ^2\right)+1\right) \cos (\xi  t)\right)-\beta  k \ln (t+\epsilon )\Bigg]\nonumber\\
&-&\Bigg|\frac{(k \ln (t+\epsilon )+t)}{4 \pi  k \ln (t+\epsilon ) (3 k \ln (t+\epsilon )+4 t)}\Bigg[48 \alpha^3 \lambda  e^{3 \zeta  t} \sin ^2(\xi  t) (3 k \ln (t+\epsilon )+4 t) (\zeta  \sin (\xi  t)\nonumber\\
&+&\xi  \cos (\xi  t))+3 \alpha ^2 \lambda  e^{2 \zeta  t} (3 k \ln (t+\epsilon )+4 t) \left(-5 \zeta ^2+\xi ^2-10 \zeta  \xi  \sin (2 \xi  t)+5 (\zeta -\xi )\right. \nonumber\\
&\times&\left.(\zeta +\xi ) \cos (2 \xi  t)\right)+\alpha  e^{\zeta  t} (3 k \ln (t+\epsilon )+4 t) \left(\zeta  \left(-6 \zeta ^2 \lambda +18 \lambda  \xi ^2+1\right) \sin (\xi  t)\right.\nonumber\\
&+&\left.\xi  \left(6 \lambda  \left(\xi ^2-3 \zeta ^2\right)+1\right) \cos (\xi  t)\right)-\beta  k \ln (t+\epsilon )\Bigg]
\Bigg|,
\end{eqnarray}

\begin{eqnarray}
\rho-3p&=&\frac{1}{4 \pi  k \log (t+\epsilon )}\Bigg[48 \alpha ^3 \lambda  e^{3 \zeta  t} \sin ^2(\xi  t) (3 k \log (t+\epsilon )+4 t) (\zeta  \sin (\xi  t)+\xi  \cos (\xi  t))\nonumber\\
&+&3 \alpha ^2 \lambda  e^{2 \zeta  t} (3 k \log (t+\epsilon )+4 t) \left(-5 \zeta ^2+\xi ^2-10 \zeta  \xi  \sin (2 \xi  t)+5 (\zeta -\xi ) (\zeta +\xi )\right.\nonumber\\
&\times& \left.\cos (2 \xi  t)\right)+\alpha  e^{\zeta  t} (3 k \log (t+\epsilon )+4 t) \left(\zeta  \left(-6 \zeta ^2 \lambda +18 \lambda  \xi ^2+1\right) \sin (\xi  t)\right.\nonumber\\
&+&\left.\xi  \left(6 \lambda  \left(\xi ^2-3 \zeta ^2\right)+1\right) \cos (\xi  t)\right)-\beta  k \log (t+\epsilon )\Bigg].
\end{eqnarray}

\noindent
\textbf{Case II:} $\omega(t)=\frac{r}{\ln t}-s$.\\

\noindent
\textbf{Subcase II(a):} $\lambda=0$.\\
\begin{eqnarray}
\rho+p&=&\frac{\ln (t)}{4 \pi }\Bigg(1+\frac{r}{\ln t}-s\Bigg)\left(-\frac{\alpha  e^{\zeta  t} (\zeta  \sin (\xi  t)+\xi  \cos (\xi  t))}{r-s \ln (t)+\ln (t)}-\frac{\beta}{-3 r+3 s \ln (t)+\ln (t)}\right),
\end{eqnarray}

\begin{eqnarray}
\rho+3p&=&\frac{\ln (t)}{4 \pi }\Bigg(1+3\left(\frac{r}{\ln t}-s\right)\Bigg)\left(-\frac{\alpha  e^{\zeta  t} (\zeta  \sin (\xi  t)+\xi  \cos (\xi  t))}{r-s \ln (t)+\ln (t)}\right.\nonumber\\
&-&\left.\frac{\beta}{-3 r+3 s \ln (t)+\ln (t)}\right),
\end{eqnarray}

\begin{eqnarray}
\rho-|p|&=&\frac{\ln (t)}{4 \pi } \left(-\frac{\alpha  e^{\zeta  t} (\zeta  \sin (\xi  t)+\xi  \cos (\xi  t))}{r-s \ln (t)+\ln (t)}-\frac{\beta}{-3 r+3 s \ln (t)+\ln (t)}\right)\nonumber\\
&-&\Bigg|\frac{\ln (t)}{4 \pi }\Bigg(\frac{r}{\ln t}-s\Bigg)\left(-\frac{\alpha  e^{\zeta  t} (\zeta  \sin (\xi  t)+\xi  \cos (\xi  t))}{r-s \ln (t)+\ln (t)}-\frac{\beta}{-3 r+3 s \ln (t)+\ln (t)}\right)\Bigg|,
\end{eqnarray}

\begin{eqnarray}
\rho-3p&=&\frac{\ln (t)}{4 \pi }\Bigg(1-3\left(\frac{r}{\ln t}-s\right)\Bigg)\left(-\frac{\alpha  e^{\zeta  t} (\zeta  \sin (\xi  t)+\xi  \cos (\xi  t))}{r-s \ln (t)+\ln (t)}\right.\nonumber\\
&-&\left.\frac{\beta}{-3 r+3 s \ln (t)+\ln (t)}\right).
\end{eqnarray}

\noindent
\textbf{Subcase II(b):} $\lambda\neq0$.\\
\begin{eqnarray}
\rho+p&=&-\frac{\log (t)\Big(1+\left(\frac{r}{\ln t}-s\right)\Big)}{4 \pi  (r-s \log (t)+\log (t)) (3 r-(3 s+1) \log (t))} \Bigg[-\beta  r+48 \alpha ^3 \zeta  \lambda  e^{3 \zeta  t} \sin ^3(\xi  t)\nonumber\\
&\times&(3 r-(3 s+1) \log (t))+\alpha  e^{\zeta  t} \sin (\xi  t) (3 r-(3 s+1) \log (t)) \left(-6 \zeta ^3 \lambda +18 \zeta  \lambda  \xi ^2+\zeta \right.\nonumber\\
&+&24 \left.\alpha ^2 \lambda  \xi  e^{2 \zeta  t} \sin (2 \xi  t)\right)-6 \alpha ^2 \lambda  \left(5 \zeta ^2-3 \xi ^2\right) e^{2 \zeta  t} \sin ^2(\xi  t) (3 r-(3 s+1) \log (t))\nonumber\\
&-&12 \alpha ^2 \lambda  \xi ^2 e^{2 \zeta  t} \cos ^2(\xi  t) (3 r-(3 s+1) \log (t))+\alpha  \xi  e^{\zeta  t} \left(-18 \zeta ^2 \lambda +6 \lambda  \xi ^2+1\right) \cos (\xi  t)\nonumber\\
&\times& (3 r-(3 s+1) \log (t))-90 \alpha ^2 \zeta  \lambda  \xi  r e^{2 \zeta  t} \sin (2 \xi  t)+90 \alpha ^2 \zeta  \lambda  \xi  s e^{2 \zeta  t} \log (t) \sin (2 \xi  t)\nonumber\\
&+&\beta  s \log (t)+30 \alpha ^2 \zeta  \lambda  \xi  e^{2 \zeta  t} \log (t) \sin (2 \xi  t)-\beta  \log (t)\Big],
\end{eqnarray}

\begin{eqnarray}
\rho+3p&=&-\frac{\log (t)\Big(1+3\left(\frac{r}{\ln t}-s\right)\Big)}{4 \pi  (r-s \log (t)+\log (t)) (3 r-(3 s+1) \log (t))} \Bigg[-\beta  r+48 \alpha ^3 \zeta  \lambda  e^{3 \zeta  t} \sin ^3(\xi  t)\nonumber\\
&\times&(3 r-(3 s+1) \log (t))+\alpha  e^{\zeta  t} \sin (\xi  t) (3 r-(3 s+1) \log (t)) \left(-6 \zeta ^3 \lambda +18 \zeta  \lambda  \xi ^2+\zeta \right.\nonumber\\
&+&24 \left.\alpha ^2 \lambda  \xi  e^{2 \zeta  t} \sin (2 \xi  t)\right)-6 \alpha ^2 \lambda  \left(5 \zeta ^2-3 \xi ^2\right) e^{2 \zeta  t} \sin ^2(\xi  t) (3 r-(3 s+1) \log (t))\nonumber\\
&-&12 \alpha ^2 \lambda  \xi ^2 e^{2 \zeta  t} \cos ^2(\xi  t) (3 r-(3 s+1) \log (t))+\alpha  \xi  e^{\zeta  t} \left(-18 \zeta ^2 \lambda +6 \lambda  \xi ^2+1\right) \cos (\xi  t)\nonumber\\
&\times& (3 r-(3 s+1) \log (t))-90 \alpha ^2 \zeta  \lambda  \xi  r e^{2 \zeta  t} \sin (2 \xi  t)+90 \alpha ^2 \zeta  \lambda  \xi  s e^{2 \zeta  t} \log (t) \sin (2 \xi  t)\nonumber\\
&+&\beta  s \log (t)+30 \alpha ^2 \zeta  \lambda  \xi  e^{2 \zeta  t} \log (t) \sin (2 \xi  t)-\beta  \log (t)\Big],
\end{eqnarray}

\begin{eqnarray}
\rho-|p|&=&-\frac{\log (t)}{4 \pi  (r-s \log (t)+\log (t)) (3 r-(3 s+1) \log (t))} \Bigg[-\beta  r+48 \alpha ^3 \zeta  \lambda  e^{3 \zeta  t} \sin ^3(\xi  t)\nonumber\\
&\times&(3 r-(3 s+1) \log (t))+\alpha  e^{\zeta  t} \sin (\xi  t) (3 r-(3 s+1) \log (t)) \left(-6 \zeta ^3 \lambda +18 \zeta  \lambda  \xi ^2+\zeta \right.\nonumber\\
&+&24 \left.\alpha ^2 \lambda  \xi  e^{2 \zeta  t} \sin (2 \xi  t)\right)-6 \alpha ^2 \lambda  \left(5 \zeta ^2-3 \xi ^2\right) e^{2 \zeta  t} \sin ^2(\xi  t) (3 r-(3 s+1) \log (t))\nonumber\\
&-&12 \alpha ^2 \lambda  \xi ^2 e^{2 \zeta  t} \cos ^2(\xi  t) (3 r-(3 s+1) \log (t))+\alpha  \xi  e^{\zeta  t} \left(-18 \zeta ^2 \lambda +6 \lambda  \xi ^2+1\right) \cos (\xi  t)\nonumber\\
&\times& (3 r-(3 s+1) \log (t))-90 \alpha ^2 \zeta  \lambda  \xi  r e^{2 \zeta  t} \sin (2 \xi  t)+90 \alpha ^2 \zeta  \lambda  \xi  s e^{2 \zeta  t} \log (t) \sin (2 \xi  t)\nonumber\\
&+&\beta  s \log (t)+30 \alpha ^2 \zeta  \lambda  \xi  e^{2 \zeta  t} \log (t) \sin (2 \xi  t)-\beta  \log (t)\Big]\nonumber\\
&-&\Bigg|\frac{\log (t)\left(\frac{r}{\ln t}-s\right)}{4 \pi  (r-s \log (t)+\log (t)) (3 r-(3 s+1) \log (t))} \Bigg[-\beta  r+48 \alpha ^3 \zeta  \lambda  e^{3 \zeta  t} \sin ^3(\xi  t)\nonumber\\
&\times&(3 r-(3 s+1) \log (t))+\alpha  e^{\zeta  t} \sin (\xi  t) (3 r-(3 s+1) \log (t)) \left(-6 \zeta ^3 \lambda +18 \zeta  \lambda  \xi ^2+\zeta \right.\nonumber\\
&+&24 \left.\alpha ^2 \lambda  \xi  e^{2 \zeta  t} \sin (2 \xi  t)\right)-6 \alpha ^2 \lambda  \left(5 \zeta ^2-3 \xi ^2\right) e^{2 \zeta  t} \sin ^2(\xi  t) (3 r-(3 s+1) \log (t))\nonumber\\
&-&12 \alpha ^2 \lambda  \xi ^2 e^{2 \zeta  t} \cos ^2(\xi  t) (3 r-(3 s+1) \log (t))+\alpha  \xi  e^{\zeta  t} \left(-18 \zeta ^2 \lambda +6 \lambda  \xi ^2+1\right) \cos (\xi  t)\nonumber\\
&\times& (3 r-(3 s+1) \log (t))-90 \alpha ^2 \zeta  \lambda  \xi  r e^{2 \zeta  t} \sin (2 \xi  t)+90 \alpha ^2 \zeta  \lambda  \xi  s e^{2 \zeta  t} \log (t) \sin (2 \xi  t)\nonumber\\
&+&\beta  s \log (t)+30 \alpha ^2 \zeta  \lambda  \xi  e^{2 \zeta  t} \log (t) \sin (2 \xi  t)-\beta  \log (t)\Big]\Bigg|,
\end{eqnarray}

\begin{eqnarray}
\rho-3p&=&-\frac{\log (t)\Big(1-3\left(\frac{r}{\ln t}-s\right)\Big)}{4 \pi  (r-s \log (t)+\log (t)) (3 r-(3 s+1) \log (t))} \Bigg[-\beta  r+48 \alpha ^3 \zeta  \lambda  e^{3 \zeta  t} \sin ^3(\xi  t)\nonumber\\
&\times&(3 r-(3 s+1) \log (t))+\alpha  e^{\zeta  t} \sin (\xi  t) (3 r-(3 s+1) \log (t)) \left(-6 \zeta ^3 \lambda +18 \zeta  \lambda  \xi ^2+\zeta \right.\nonumber\\
&+&24 \left.\alpha ^2 \lambda  \xi  e^{2 \zeta  t} \sin (2 \xi  t)\right)-6 \alpha ^2 \lambda  \left(5 \zeta ^2-3 \xi ^2\right) e^{2 \zeta  t} \sin ^2(\xi  t) (3 r-(3 s+1) \log (t))\nonumber\\
&-&12 \alpha ^2 \lambda  \xi ^2 e^{2 \zeta  t} \cos ^2(\xi  t) (3 r-(3 s+1) \log (t))+\alpha  \xi  e^{\zeta  t} \left(-18 \zeta ^2 \lambda +6 \lambda  \xi ^2+1\right) \cos (\xi  t)\nonumber\\
&\times& (3 r-(3 s+1) \log (t))-90 \alpha ^2 \zeta  \lambda  \xi  r e^{2 \zeta  t} \sin (2 \xi  t)+90 \alpha ^2 \zeta  \lambda  \xi  s e^{2 \zeta  t} \log (t) \sin (2 \xi  t)\nonumber\\
&+&\beta  s \log (t)+30 \alpha ^2 \zeta  \lambda  \xi  e^{2 \zeta  t} \log (t) \sin (2 \xi  t)-\beta  \log (t)\Big].
\end{eqnarray}

\end{document}